\documentclass[aps,10pt,twocolumn,superscriptaddress]{revtex4}

\usepackage{amssymb,amsmath}
\usepackage{subfig}
\usepackage{graphicx}
\usepackage{color}
\usepackage{hyperref}
\usepackage{listings}
\usepackage{ragged2e}
\usepackage{mhchem}
\usepackage{tabu}

\hypersetup{
    bookmarks=true,         
    unicode=false,          
    pdftoolbar=true,        
    pdfmenubar=true,        
    pdffitwindow=false,     
    pdfstartview={FitH},    
    pdfauthor={Armaos, Badounas, Deligiannis},     
    colorlinks=true,       
    linkcolor=blue,          
    citecolor=blue,        
}
\graphicspath{{figures/}}

\begin{document}

\definecolor{dkgreen}{rgb}{0,0.6,0}
\definecolor{gray}{rgb}{0.5,0.5,0.5}
\definecolor{mauve}{rgb}{0.58,0,0.82}

\captionsetup{justification=justified,singlelinecheck=false,labelfont=large}

\lstset{frame=tb,
  	language=Matlab,
  	aboveskip=3mm,
  	belowskip=3mm,
  	showstringspaces=false,
  	columns=flexible,
  	basicstyle={\small\ttfamily},
  	numbers=none,
  	numberstyle=\tiny\color{gray},
 	keywordstyle=\color{blue},
	commentstyle=\color{dkgreen},
  	stringstyle=\color{mauve},
  	breaklines=true,
  	breakatwhitespace=true
  	tabsize=3
}

\title{Computational Chemistry on Quantum Computers: Ground state estimation}

\author{V. Armaos}
\email[Corresponding author: ]{billy.armaos@gmail.com}
\affiliation{Pi-dust R\&D, Greece}
\affiliation{Laboratory of Atmospheric Physics, Department of Physics, University of Patras, Greece}

\author{Dimitrios A. Badounas}
\email[Corresponding author: ]{badounasdimitris@gmail.com}
\affiliation{Pi-dust R\&D, Greece}
\affiliation{Department of Material Science, University of Patras, Greece}

\author{Paraskevas Deligiannis}
\affiliation{Pi-dust R\&D, Greece}

\date{\today}

\begin{abstract}
    We present computational chemistry data for small molecules ($CO$, $HCl$, $F_2$, $NH_4^+$, $CH_4$, $NH_{3}$, $H_3O^+$, $H{_2}O$, $BeH_{2}$, $LiH$, $OH^-$, $HF$, $HeH^+$, $H_2$), obtained by implementing the Unitary Coupled Cluster method with Single and Double excitations (UCCSD) on a quantum computer simulator. We have used the Variational Quantum Eigensolver (VQE) algorithm to extract the ground state energies of these molecules. This energy data represents the expected ground state energy that a quantum computer will produce for the given molecules, on the STO-3G basis. Since there is a lot of interest in the implementation of UCCSD on quantum computers, we hope that our work will serve as a benchmark for future experimental implementations.
  
\end{abstract}

\maketitle

\section{\label{Introduction}Introduction}
    The most natural application of quantum computers is to simulate quantum mechanical systems\cite{Feynman1986}. The appearance of quantum algorithms\cite{3,4,5} and subsequently of quantum processors\cite{2018APS..MARA33001K,1905.11349} has made this argument solid enabling quantum computing. Several important problems which are traditionally hard to solve on classical computers can be now addressed on their quantum counterparts\cite{6,7}.
    
    Quantum chemistry  is one of the most promising applications of quantum computing. With only a few hundreds of qubits, quantum computers seem to outperform classical computers in the determination of molecular energies within chemical  accuracy\cite{8}. Early quantum computation algorithms have been used to estimate the energy of small molecules, however simulations of larger molecules still remain out of reach because of the limitation in the number of qubits and coherence times in noisy intermediate scale quantum (NISQ) devices.
    
    Sophisticated methods have been used to reduce the cost of quantum chemistry simulations\cite{32,33,34}, such as Hybrid Quantum Classical (HQC) algorithms\cite{19}. Here we focus on one of these HQC algorithms, the Variational Quantum eigensolver (VQE). In this algorithm, the computation is split into several quantum sub-tasks. A classical optimizer controls the experiments performed on the quantum computer to determine the parameters that minimize the expectation value of the Hamiltonian. This is equivalent to finding the eigenvector of the Hamiltonian with the smallest eigenvalue, hence the name of the method.

    In order to solve quantum chemistry problems on quantum computers we are using the Unitary Couple Cluster (UCC) method. The benefit of UCC over classical computational chemistry methods is its ability to take into account both static and dynamic electron correlation. In this work we are using the UCC method with Single and Double excitations (UCCSD) on the Qiskit statevector simulator in order to estimate the ground state energy of small molecules ($CO$, $HCl$, $F_2$, $NH_4^+$, $CH_4$, $NH_{3}$, $H_3O^+$, $H{_2}O$, $BeH_{2}$, $LiH$, $OH^-$, $HF$, $HeH^+$, $H_2$) within chemical accuracy. Our results confirm the viability of UCCSD for quantum chemistry on quantum computers while providing data to compare against, for future UCCSD calculations.

\section{\label{Framework}Theoretical Framework}
    In this section we will present all the steps necessary for performing computational chemistry simulations on quantum computers. We start from the second quantization and the introduction of Slater determinants. We discuss the ways to encode information to quantum computers and finally, we talk about VQE and the specifics of our implementation.

    \subsection{Second Quantization}
        We begin our analysis from Schrodinger's Equation $\hat{H}|\Psi\rangle=E|\Psi\rangle$. We want to solve this for an arbitrary molecule. The general molecular Hamiltonian describes a system of $N_e$ electrons and $N_n$ nuclei in a composite potential comprised by the Coulomb potential that each particle produces. However, to simplify this problem it is common to assume the Born-Oppenheimer approximation where the nuclei are treated as stationary points in space. This way the Hamiltonian is written in atomic units as
        
        \begin{equation} \label{ham_BO_atomic}
            \hat{H} = -\sum_{i}\frac{\nabla^2_i}{2}
                -\sum_{i,I}\frac{Z_I}{|\pmb{r}_i-\pmb{R}_I|}
                +\sum_{i\ne j}\frac{1}{2|\pmb{r}_i-\pmb{r}_j|},
        \end{equation}
        
        \noindent where, the indices $i,j$ refer to electrons and $I$ to nuclei. $Z_I$ is the atomic number of the $I^{th}$ nucleus, $\pmb{R}_I$ is the fixed location of the $I^{th}$ nucleus and the $\pmb{r}_{i}$'s are the variables of the Hamiltonian, representing the position of the electrons.
        Note that the first term describes the kinetic energy of the electrons, the second the Coulomb interaction between the electrons and the fixed nuclei and the third the Coulomb interaction between different electrons.
        
        At this point we introduce the second quantization. This comes with the anti-commuting creation and annihilation operators $\alpha^\dagger$ and $\alpha$ that satisfy
        
        \begin{equation} \label{anticommutation_relations}
            \begin{aligned}
                {}
                & \{\alpha_p,\alpha_q^\dagger\}=\delta_{pq} \\
                & \{\alpha_p,\alpha_q\}=0 \\
                & \{\alpha_p^\dagger,\alpha_q^\dagger\}=0.
            \end{aligned}
        \end{equation}
        
        In the second quantization framework the Hamiltonian of equation \ref{ham_BO_atomic} is written as
        
        \begin{equation} \label{ham_second_quant}
            \hat{H} = \sum_{pq}h_{pq}\alpha_{p}^\dagger\alpha_{q}+
                \frac{1}{2}\sum_{pqrs}h_{pqrs}\alpha_{p}^\dagger\alpha_{q}^\dagger\alpha_{r}\alpha_{s},
        \end{equation}
    
        \noindent where 
        
        \begin{equation} \label{integrals}
            \begin{aligned}
                {}
                & h_{pq}=\int d\pmb{x}\phi_{p}^\ast(\pmb{x})\Bigg(-\frac{\nabla^2_i}{2} - \sum_{I}\frac{Z_I}{|\pmb{R}_I - \pmb{r}_i|}\Bigg)\phi_q(\pmb{x}) \\
                & h_{pqrs}=\int d\pmb{x}_1d\pmb{x}_2\frac{\phi^*_p(\pmb{x}_1)\phi^*_q(\pmb{x}_2)\phi_s(\pmb{x}_1)\phi_r(\pmb{x}_2)}{|\pmb{x}_1 - \pmb{x}_2|}.
            \end{aligned}
        \end{equation}
        
        The $\phi(\pmb{x})$ are the wavefunctions (orbitals) of the basis into which we chose to project the Hamiltonian. The choice of a basis set is crucial for computational chemistry. As a rule, the bigger the basis, the more accurately it describes the Hamiltonian. However, having a bigger basis makes the computation harder. In our case, we restrict ourselves to the STO-3G basis, which is small but it allows us to perform calculations on the small amount of qubits that we have available.
    
    \subsection{\label{Encoding}Encoding to Quantum Computers}

        Although the process discussed in the previous section is inherently quantum mechanical, all this formalism is purely theoretical in the sense that there is yet no mention of computing either classical or quantum. Indeed, in order to start extracting useful information from the theory that we have discussed so far, we need a way to encode information in terms that can be applied on a computational machine. Here, the discussion is focused solely on quantum computers.
        
        We need a way of mapping a fermionic state (Slater Determinant) of the form $|f_{M-1},...f_i,...f_1,f_0\rangle$ (where $f_i=0,1$) to a qubit state of a similar form $|q_{M-1},...q_i,...q_1,q_0\rangle$ (where $q_i=0,1$). Choosing a map for which $q_i=f_i, \forall \,i \in [0,M-1]$ is very compelling, since this way the $q_i$'s have a clear physical interpretation (occupied or unoccupied orbital). However, once we try to map the fermionic creation and annihilation operators to qubits, we realise that they do not correspond to simple raising and lowering operators ($\sigma^+$ and $\sigma^-$ respectively).
        
        Of course, that's to be expected. Electrons are indistinguishable fermions, while qubits behave as distinguishable particles. Therefore, they obey different statistics.
        Thankfully, there is still a way to use this intuitive mapping, if we define
        
        \begin{equation}
            \begin{aligned}
                {}
                & \alpha_{p}^\dagger = (\prod_{m<p}\sigma_m^z)\cdot \sigma_p^+ \\
                & \alpha_{p} = (\prod_{m<p}\sigma_m^z)\cdot \sigma_p^-.
            \end{aligned}
        \end{equation}
        
        In this process we have recovered the Jordan-Wigner transformation\cite{17,18}, which has the benefit of simplicity but comes with some drawbacks. Most notably, it uses more qubits than other alternatives (e.g. the Bravyi-Kitaev transformation\cite{25,26}).
        
        For our purposes, intuitiveness was important, so we decided to use the Jordan-Wigner transformation despite its drawbacks.
        
    \subsection{Variational Quantum Eigensolver}
    
    \begin{figure}[t]
        \includegraphics[width=1.0\linewidth]{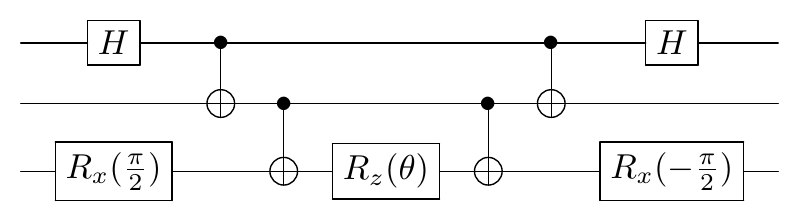}
        
        \captionsetup{width=.9\linewidth,justification=centerlast,singlelinecheck=off}
        
        \caption{\small\textbf{Circuit design for quantum chemistry.} Circuit implementation of the UCC operator $e^{-i\theta XZY}$ for 3 qubits. $\theta$ controls the amplitude of the excitation. The $H$ and $R_z(\frac{\pi}{2})$ gates facilitate a basis change, so that the applied operator to the selected qubit is $X$ and $Y$ respectively, instead of $Z$. Serves as an essential circuit for any type of UCC simulation.\cite{30}}
        \vspace{-0.2cm}
        \rule{\linewidth}{0.5pt}

        \label{fig:cirquit}
    \end{figure}

        Now that we have a way of encoding the computational chemistry problem to quantum computers, it's time to discuss the algorithm used to solve it. That is the VQE algorithm, which is a hybrid quantum-classical process.
        
        The cornerstone of VQE is the Rayleigh-Ritz variational principle
        \begin{equation} \label{Reyleigh-Ritz}
            \langle\Psi|\hat{H}|\Psi\rangle\geq E_0,
        \end{equation}
        \noindent that states the expectation value of a Hamiltonian will always be greater than or equal to its smallest eigenvalue $E_{0}$, for any wavefunction $|\Psi\rangle$. The equality holds true only when $|\Psi\rangle$ is the corresponding eigenvector of $E_{0}$.
        
        In our case, we parametrize the wavefunction as $|\Psi(\pmb{\theta})\rangle$, where $\pmb{\theta} = (\theta_0, \theta_1, \theta_2,...)$ and $\theta_i$ is the parameter of the $i^{th}$ excitation. More info about the VQE implementation can be found at McArdle et al.\cite{30}. Our specific implementation is outlined in the following steps:
        
        \begin{enumerate}
            \item An ab-initio computational chemistry program is used (in our case Psi4) to perform a (classically tractable) Hartree-Fock (H-F) geometry optimization calculation on the chosen molecule for the selected computational chemistry basis (e.g. STO-3G, 6-31G ,cc-pVDZ). The resulting optimal geometry is fed to the next step of the algorithm.
            
            \item The H-F optimal geometry is used as the ansatz for the UCCSD geometry optimization algorithm. Assuming that the H-F geometry is sufficiently close to the UCCSD optimal geometry, we fit a paraboloid to find the actual minimum. See \autoref{contour}.
            
            \item For each distinct geometry,  we run again a classical H-F calculation to obtain the fermionic Hamiltonian and the ground state energy of the configuration.
            
            \item We use the Jordan-Wigner encoding to transform the fermionic Hamiltonian in to the corresponding qubit Hamiltonian and perform the following VQE algorithm.
            
            \item We initialize the quantum computer at the encoded H-F ground state, $|00...0011...11\rangle$ ($N$ ones and $M-N$ zeros), where $N$ is the number of electrons and $M$ the number of orbitals considered. Using circuits similar to the one pictured in \autoref{fig:cirquit}, we perform double and single excitations on the initial state. We write the resulting state as $|\Psi (\pmb{\theta})\rangle$, where $\pmb{\theta} = (\theta_0, \theta_1, \theta_2,...)$ and $\theta_i$ is the parameter of the $i^{th}$ excitation.
            
            \item A classical optimizer is used to find the configuration $\pmb{\theta}_{min}$ that minimizes the energy of the specific geometry. The corresponding state $|\Psi (\pmb{\theta}_{min})\rangle$ is the UCCSD ground state for the specific geometry and the expectation value of the Hamiltonian $\langle\Psi(\pmb{\theta}_{min})|\hat{H}|\Psi(\pmb{\theta}_{min})\rangle$ is the corresponding  ground state energy.
            
            \item The energy of any $|\Psi (\pmb{\theta})\rangle$ can be found by measuring the expectation values of the operators that comprise the Hamiltonian (McArdle et al.\cite{30}). Since we are using a statevector simulator, we have the benefit of retrieving the expectation value directly from the wavefunction. In an actual quantum computer, we would have to execute the same experiment multiple times to evaluate the expectation value of each operator. 
            
            \item In our case, in order to perform the classical optimization part of the algorithm, we just iterate through the excitations and optimize each one sequentially until the energy converges. We found that in order to achieve chemical accuracy, three iterations through the list of excitations are usually sufficient. 
        \end{enumerate}
    
    \begin{figure*}[t]
    \centering
    \includegraphics[width=1.0\textwidth]{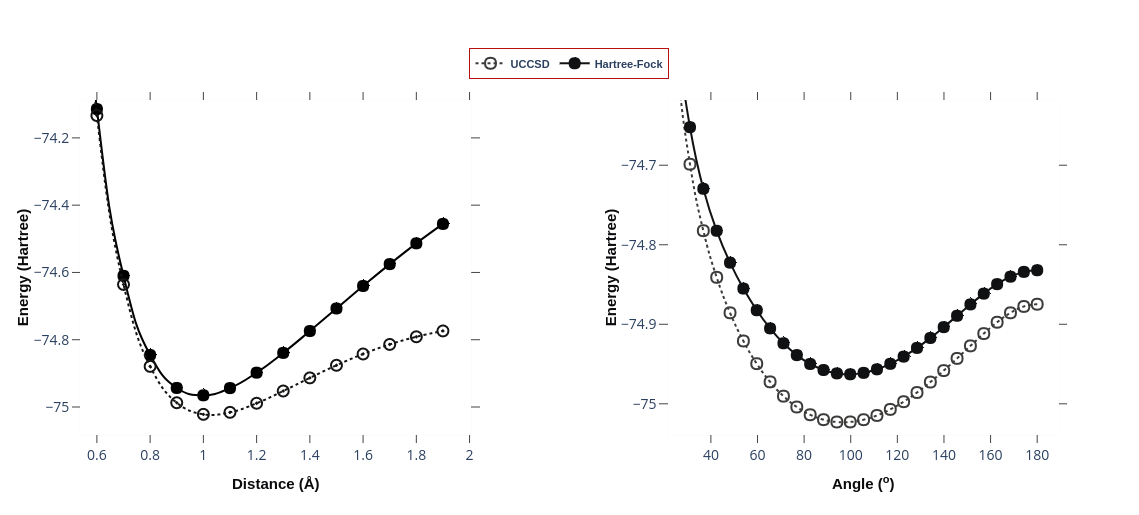}
    
    \captionsetup{width=1.0\textwidth,justification=centerlast,singlelinecheck=on}
    
    \caption{\textbf{Illustration of geometry parameters of $\boldsymbol{H_2O}$.} On the left panel, we present the ground state energy of $H_2O$ as a function of the distance between the oxygen atom and any of the two Hydrogen atoms. On the right panel, the energy is a function of the angle between the two $OH$ bonds. In each graph we keep constant one of the geometry variables. The empty circles correspond to the UCCSD ground state energy while the filled in dots to the Hartree-Fock energy.}
    \vspace{-0.2cm}
    \rule{\linewidth}{0.5pt}

\label{fig:single_param}
\end{figure*}

    
        
    
\section{\label{Results}Results}

\begin{table}[t]
    \resizebox{1.0\linewidth}{!}{%
    \begin{ruledtabular}
        \begin{tabular}{ccccc}

            Molecule & Qubits & Energy & Distance & Angle \\
            & & \scriptsize{(Hartree)} & \scriptsize{(\AA)} & \scriptsize{($^{\circ}$)} \\
            \hline
            
            $H_2$       & 4  & -1.137306   & 0.735 & --    \\
            $HeH^+$     & 4  & -2.862695   & 0.913 & --    \\
            $LiH$       & 12 & -7.882752   & 1.546 & --    \\ 
            $OH^-$      & 12 & -74.095341  & 1.112 & --    \\ 
            $HF$        & 12 & -98.603302  & 0.995 & --    \\ 
            $BeH_2$     & 14 & -15.594875  & 1.316 & --    \\ 
            $H_2O$      & 14 & -75.023189  & 1.028 & 96.9  \\ 
            $H_3O^+$    & 16 & -75.396782  & 1.021 & 69.0  \\ 
            $NH_3$      & 16 & -55.528054  & 1.070 & 62.2  \\ 
            $CH_4$      & 18 & -39.806790  & 1.108 & --    \\ 
            $NH_4^+$    & 18 & -55.954449  & 1.067 & --    \\
            $F_2$       & 20 & -196.050161 & 1.387 & --    \\
            $HCl$       & 20 & -455.157067 & 1.342 & --    \\
            $CO$        & 20 & -111.363038 & 1.182 & --    \\

        \end{tabular}
    \end{ruledtabular}
    }
    
    \captionsetup{width=.9\linewidth,justification=centerlast,singlelinecheck=off}    
    
    \caption{\small \label{tab:geometries}\textbf{UCCSD Ground State Energy and Geometries of small Molecules.} For diatomic molecules the only optimized variable is the distance between the two molecules. Same is true for $BeH_2$, where we assumed that the three atoms are colinear and for $CH_4$ and $NH_4^+$ where we assumed tetrahedral geometries. In these cases, the distance shown is the distance between the heaviest atom and any hydrogen. For $H_2O$, we provide the angle between the two $OH$ bonds. In the case of $NH_3$ and $H_3O^+$, we provide the angle between any $NH$ or $OH$ bond respectively and the symmetry axis of the molecule.}
    \vspace{-0.2cm}
    \rule{\linewidth}{0.5pt}
    
\end{table}

    We carried out quantum chemistry calculations of small molecules ($CO$, $HCl$, $F_2$, $NH_4^+$, $CH_4$, $NH_{3}$, $H_3O^+$, $H{_2}O$, $BeH_{2}$, $LiH$, $OH^-$, $HF$, $HeH^+$, $H_2$). These calculations were executed on the Qiskit state vector simulator\cite{29}. We managed to push our hardware to use up to 20 qubits in a reasonable amount of time, using a reasonable amount of memory. In order to map the fermionic Hamiltonian into a qubit Hamiltonian we've used the Jordan-Wigner encoding, as discussed in section \ref{Encoding}. We've used the minimal basis STO-3G (Slater  Type  Orbitals, where each Slater orbital is approximated by fitting 3  Gaussian orbitals). This basis is not common in practical ab-initio calculations since it cannot provide very accurate results, however we had to use it since it requires the smallest possible number of qubits. We could reduce the amount of orbitals considered by freezing the inner-core electrons, however, this way we would loose some of the small energy contributions of dynamic correlation, which is the main benefit of this method.
    
    In \autoref{tab:geometries}, we present the main results of our simulations. One can see the simulated molecules, the amount of qubits used for simulating each one of them, their ground state energy and their optimal geometry. The amount of qubits used was always equal to the number of spin-orbitals of the STO-3G basis. It is possible to simulate some of these molecules using fewer qubits by using a different encoding, by exploiting symmetries of the molecule, or by freezing the core orbitals. Equivalently, it is possible to harness these qubit reduction techniques to simulate bigger molecules while staying in the limit of 20 qubits. However, that was not a priority for this study.
    
\begin{table}[!h]
    \resizebox{1.0\linewidth}{!}{%
    \begin{ruledtabular}
        \begin{tabular}{cccc}

            Molecule & UCCSD & CCSD & Experimental \\
            \hline
            
            $H_2$       & -1.137306   & -1.137306   & -1.1371\cite{27}          \\
            $HeH^+$     & -2.862695   & -2.862695   & -2.865$\pm$0.008\cite{20} \\
            $LiH$       & -7.882752   & -7.882752   & -7.8807\cite{27}          \\
            $OH^-$      & -74.095341  & -73.968894  & --                        \\
            $HF$        & -98.603302  & -98.603302  & --                        \\
            $BeH_2$     & -15.594875  & -15.594861  & $\simeq$ -15.4\cite{23}     \\
            $H_2O$      & -75.023189  & -75.023141  & -74.985\cite{21}          \\
            $H_3O^+$    & -75.396782  & -75.396854  & --                        \\
            $NH_3$      & -55.528054  & -55.527975  & --                        \\
            $CH_4$      & -39.806790  & -39.806743  & --                        \\
            $NH_4^+$    & -55.954449  & -55.954426  & --                        \\
            $F_2$       & -196.050161 & -196.050162 & --                        \\
            $HCl$       & -455.157067 & -455.157068 & --                        \\
            $CO$        & -111.363038 & -111.362661 & --                        \\

        \end{tabular}
    \end{ruledtabular}
    }
    
    \captionsetup{width=.9\linewidth,justification=centerlast,singlelinecheck=off}
    
    \caption{\small \label{tab:Experimental}\textbf{Comparison of UCCSD, CCSD and experimental UCCSD ground state energies of small molecules.} In the "UCCSD" column we present the ground state energy that we obtained from our calculations (duplicate of "Energy" column of \autoref{tab:geometries}). The "CCSD" column shows the ground state energy of the classical CCSD method produced by Psi4. The "Experimental" column corresponds to experimental data obtained by UCCSD calculations on quantum computers instead of quantum computer simulators. All energies are given in Hartrees.}
    \vspace{-0.2cm}
    \rule{\linewidth}{0.5pt}
    
\end{table}
    
    \subsection{Comparison with Classical and Experimental results}
        The main purpose of this study was to understand more about the UCCSD method, as a way to simulate molecules. To do so, we wanted to compare our results with the classical bibliography. The ideal benchmark would be to produce classical UCCSD data to compare against. However, UCCSD is not well suited for classical computers and it is therefore difficult to retrieve this kind of data. We decided to instead compare with CCSD. Both methods try to solve the same problem in a similar manner, while producing slightly different results in the limit of a small basis set. The CCSD data were obtained by running the classical computational chemistry package Psi4 for each molecule. 
    
        Another approach for benchmarking our code, was to compare it to UCCSD data from actual quantum computers. The experimental data, however, is obtained from NISQ computers and therefore it is not precise enough for a thorough comparison.
    
        Regardless, we compiled \autoref{tab:Experimental}, where we compare our results with CCSD and experimental UCCSD data. 

    \subsection{Water Energy Curves}
    
\begin{figure}[t]
    \includegraphics[width=.98\linewidth]{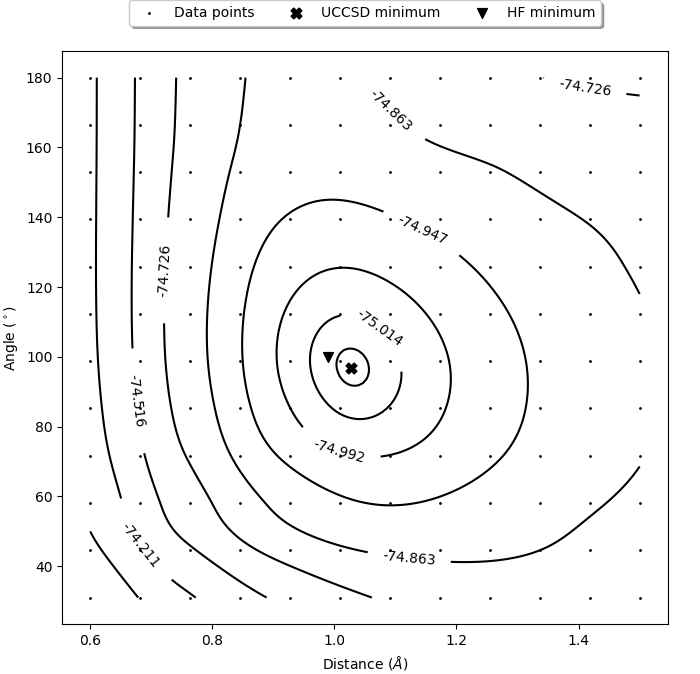}
    
    \captionsetup{width=.9\linewidth,justification=centerlast,singlelinecheck=off}
    
    \caption{\small \textbf{$\boldsymbol{H_2O}$ Geometry--Energy graph.} Contour plot of UCCSD ground state energy of $H_2O$ using VQE. The small dots represent the location of the actual data points from which we produced this plot. The bold cross overlaid in the center corresponds to the actual energy minimum of the molecule and was found independently, by running the geometry optimization algorithm. The corresponding energy value can be found on \autoref{tab:geometries}. The triangle close to the minimum is the Hartree-Fock optimal geometry, which is the initial ansatz to our UCCSD energy optimization algorithm. Notice that the energy scale of the contours is not linear, in order to depict more clearly the substantial variability of the energy scale over the span of the graph. The innermost contour is too small to overlay the label of its energy, which is -75.022. All energies are given in Hartrees.}
    \vspace{-0.2cm}
    \rule{\linewidth}{0.5pt}
    
    \label{contour}
\end{figure}     

        We have now presented the most important part of our results. Here, we would like to delve into the specifics of our methodology. In the following graphs we present the UCCSD ground state energy dependence on the geometry of a molecule. We chose the $H_{2}O$ molecule for our graphs because it is sufficiently fast to obtain data (requires only 12 qubits), while it remains sufficiently complicated (2 geometrical degrees of freedom, distance and angle).
        
        In \autoref{contour}, we present the energy dependence on the geometry of the molecule. Since we have two geometry variables, the corresponding graph was chosen to be a contour plot.
        
        To more clearly depict the energy dependence on each of the two variables, we include two additional graphs in \autoref{fig:single_param}. Each one shows the dependence of the energy on one variable, while keeping the other variable constant and equal to the value where the minimum is presented. Specifically, the energy-distance graph is for an angle of $96.9^{\circ}$ and the energy-angle graph is for a distance of $1.028\textrm{\AA}$.
        
\begin{figure}[!t]
    \includegraphics[width=1\linewidth]{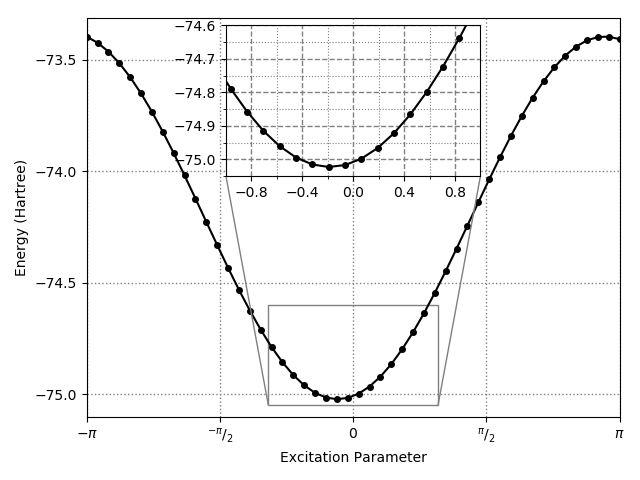}
    
    \captionsetup{width=.9\linewidth,justification=centerlast,singlelinecheck=off}
    
    \caption{\small \textbf{Double Excitation Contribution to Energy.} In this plot we present the energy contribution of the $e^{\theta(\alpha^\dagger_{13}\alpha^\dagger_{12}\alpha_5\alpha_4-\alpha^\dagger_4\alpha^\dagger_5\alpha_{12}\alpha_{13})}$ excitation. For $\theta=0$, the excitation has no effect on the state and the energy. For $\theta=\pm\pi$, the excitation takes full effect, equivalent to acting with the $\alpha^\dagger_{13}\alpha^\dagger_{12}\alpha_5\alpha_4$ operator on the state. As it is expected, the minimum lies close to $0$, since the double excitations have a small (though significant) effect on the energy compared to the Hartree-Fock state. In this case the minimum is lies at $-0.184$}
    \vspace{-0.2cm}
    \rule{\linewidth}{0.5pt}

    \label{excitations}
\end{figure}

        Finally, in \autoref{excitations}, we present the energy dependence on the excitation parameter. More specifically, we found the most significant excitation, in terms  of its contribution to the ground state. For this excitation, we varied the excitation parameter from $-\pi$ to $\pi$ and evaluated the energy corresponding to it. Note that the minimum lies close to $0$. This means that the excitation contributes only by a small factor to the ground state. This is expected, since the main Slatter determinant has to be the Hartree-Fock state, while double excitations present small corrections to the ground state.

\section{\small \label{Conclusions}Conclusions \& Future Work}

    We have now presented the arguments supporting that quantum computers are well suited for computational chemistry calculations. We described the theoretical framework that enables the application of quantum computational chemistry, from Schrodinger's equation to Slater determinants and from encoding methods all the way to VQE. We have also described the algorithm that we used to obtain our results that we have listed in \autoref{tab:geometries}.
    
    By comparing our UCCSD calculations with the classical CCSD ones (\autoref{tab:Experimental}), the argument for computational chemistry on quantum computers becomes clear. UCCSD is a superior method, capable of taking into account both static and dynamic electron correlations in molecular systems. However, much more work needs to be done until we have reached an era where quantum computers are able to outperform classical ones in computational chemistry calculations.
    
    An important factor that limits our ability to simulate bigger molecules is the encoding method that we used. The Jordan-Wigner encoding, although simple and intuitive, requires more qubits to simulate the same systems than other alternatives e.g. the Bravyi-Kitaev encoding. Nevertheless, the Jordan-Wigner encoding served us well in achieving our goal, which was to provide data for enough molecules to facilitate the benchmarking of future quantum computing implementations of UCCSD.
    
    Finally, we would like to add that we are dedicated to expand the list of molecules that we have simulated and the configurations that we have used. Future work will include different small molecules, basis sets, encodings, or sets of excitations (e.g. UCCSDT or UCCD).

 \section{\label{acknowledgement}Acknowledgements}
    Here, we would like to thank Prof. Athanassios Argiriou, Georgios Papasotiropoulos (MSc), Theofilos Sotiropoulos-Michalakakos (MPhil) and Dr. Christos Garoufalis for useful insights on this work. We would also like to thank the Laboratory of Atmospheric Physics of the University of Patras and "DiscoverYourWay" for their support.

\bibliography{bib}
\end{document}